**Dipole moments from atomic-number-dependent potentials in analytic density-functional theory**

Brett I. Dunlap,

Code 6189, Theoretical Chemistry Section, US Naval Research Laboratory, Washington, DC 20375-5342

Shashi P. Karna

US Army Research Laboratory, Weapons and Materials Research Directorate, ATTN: AMSRD-ARL-WM, Bldg. 4600, Aberdeen Proving Ground, MD 21005-5069

Rajendra R. Zope

Department of Physics, University of Texas at El Paso, El Paso, TX 79959

**Abstract**

Molecular dipole moments of analytic density-functional theory (ADFT) are investigated. The effect of element-dependent exchange potential on these moments are examined by comparison with conventional quantum-chemical methods and experiment for the subset of the extended G2 set of molecules that have nonzero dipole moment. Fitting the Kohn-Sham potential itself makes a mean absolute error of less than 0.1 Debye. Variation of $\alpha$ (Slater's exchange parameter) values has far less effect on dipole moments than on energies. It is argued that in variable $\alpha$ methods one should choose the smaller of the two rather than the geometric mean of the two $\alpha$ values for the heteroatomic part of the linear-combination-atomic-orbitals (LCAO) density. Calculations on the dipole moment of $NH_2(CH)_{24}NO_2$ are consistent with earlier calculations and show that varying the differences between $\alpha$ values for atoms with different atomic numbers has only short-ranged electrostatic effects.

## I. Introduction

Molecular density-functional theory (DFT) [1] and *ab initio* quantum chemistry [2] differ. Broadly speaking DFT methods require numerical integration, which is their bottleneck, while Hartree-Fock (HF) -based methods do not, instead requiring analytic four-center two-electron integrals, which are their bottleneck. The completely numerical approach to DFT is well established and even practical for small molecules [3]. Approximating orbitals as a linear combination of atomic orbitals (LCAO) is more generally useful because the possible variational space of molecular wavefunctions is sharply reduced [4]. Using Gaussians in *ab initio* quantum chemistry has the added advantage of precise analytic integrals [5]. This tremendous reduction in variational freedom becomes useful only through understanding of what types of orbital variational freedom are essential for describing various chemical processes [2]. A good general-purpose orbital basis set for chemical problems today has a single basis function, which is a fixed contraction of fewer than ten primitive Gaussians of the correct atomic angular momentum for each core orbital. The valence orbitals are often split into two or three parts, respectively, double-zeta (DZ) and triple-zeta (TZ), which together contain roughly the same number of primitive Gaussians as the corresponding core orbital. One also routinely uses polarized basis sets, DZP and TZP, which include a complete shell of atomic orbitals that have angular momentum one unit higher than is occupied in the atom. Thus each angular momentum component about any atom of each non-core molecular orbital has one, two or three degrees of variational freedom. Of course, considering all angular momentum components about all atoms the dimension, N, of the variational freedom of a molecular orbital for a reasonably sized molecule still can become quite large. The size of the primitive basis is roughly twice as large, ~2N.

In Hartree's simplest self-consistent-field (SCF) model of electronic structure the electronic wavefunction is a product of molecular orbitals. The square of each orbital gives its charge distribution. The total electron-electron interaction is expressed in the prescription of DFT as the self-Coulomb energy of the entire electronic density less the self-Coulomb energy of each molecular orbital. The sophisticated *ab initio* electronic-structure calculations of today consider (multiple) Slater determinants rather than product functions and additional terms in the Hamiltonian. Thus all *ab initio* SCF calculations scale as $N^4$ and higher and are extremely challenging for large molecules and materials using DZP and TZP basis sets.

If the LCAO approximation is good for fitting the molecular orbitals, then it is good for fitting the Kohn-Sham (KS) potential [6]. Such a method scales as $N^3$ [7], but the fit of the KS potential must be as good as the fit to the molecular orbitals, which are variationally determined. Variational fitting treats the analytic KS potential on an equal footing with the analytic orbitals [8].

The central field approximation is the starting point of every numerical description of atoms [9]. In this approximation the SCF is spherically symmetric and the atomic orbitals are eigenstates of angular momentum. Thus it is a good approximation to fit the KS potential to overlapping *s*-type functions about each atom [7]. One cannot go one step further, though, and muffin-tin the KS potential, which is to fit it to spherically-symmetric form in touching (or overlapping) spheres about each atom and to a constant in the region between spheres. Such fitted KS potentials are discontinuous and thus cannot be associated with a meaningful DFT energy. If one does not have a good energy, then the essence of quantum chemistry, the variational principle, cannot be used. If DFT, which is in principle exact, were compatible with a muffin-tin KS potential then unique variational quantum chemistry would be possible. As the bond distances in any molecule grow the muffin-tin spheres grow ultimately giving an exact description of each atom in the molecule as the entire wavefunction becomes completely contained in atomic spheres. No other quantum chemistry method is both practical for molecules and precise for dissociated atoms. That was the promise of the Xα Scattered-Wave method [10]. In that Xα implementation, each atom in its muffin-tin sphere can have its own parameter α that scales Slater's density-functional expression for the average electronic exchange energy [11],

$$E_x[\rho] = -\tfrac{3}{4}\left(\tfrac{3}{\pi}\right)^{1/3}\left\langle \rho^{4/3}\right\rangle \equiv -\kappa\left\langle \rho^{4/3}\right\rangle, \tag{1}$$

where $\rho(\mathbf{r})$ is the charge density and angular braces indicate integration over all space, towards the variational value, which is two thirds as big [6, 12]. This is the non spin-polarized expression. For spin-polarized exchange energies, this expression is repeated for each spin, with $\kappa_\uparrow = \kappa_\downarrow = \kappa 2^{1/3}$, and one obtains the energy of Eq. (1) if both spin densities converge to identical densities.

It is possible to combine an LCAO treatment of the KS potential and associate different α values with different atoms in a method that has a continuous KS potential and is variational. This economical approach to electronic structure is analytic density-functional theory (ADFT) [13-15]. It requires no numerical grid at all. Its variational parameters are 10's of LCAO coefficients rather than 100's of plane-wave coefficients or numerical values at 1000's of points per atom per molecular orbital. Therefore matrix elements and thus the total energy can be computed to machine precision. This approach enables calculations on the largest molecules that be treated using an all-electron TZP basis set [16]. Its functional form is restricted but its space of atomic parameters is rich. The potential of ADFT can be assessed through comparison to experiment and other theoretical methods.

If atoms of different elements have different $\alpha$ values that extra dimension will not only affect the binding characteristics of molecules, but it will also affect the distribution of electrons within heteronuclear molecules. This will have perhaps its largest affect on dipole moments. The exchange energy, Eq. (1), is negative. If the $\alpha$ value of an atom is increased electrons lower their energy by moving onto that atom. Similarly, decreasing its $\alpha$ value will drive electrons away from an atom. The dipole moments might be so sensitive as to make the added dimension of element-dependent exchange in ADFT of no practical use. This work tests that possibility. While properties, including the dipole moment, are better if a bigger, more diffuse basis set is used for their computation than is used for geometry optimization [17], basis-set limit dipole moments [18] are not required for this assessment. The analysis is simpler if a single basis set is used for everything, particularly so, given our need for three basis sets in addition to the orbital basis set.

ADFT is only possible through variational fitting. The next section reviews variational fits of molecular Coulomb repulsion and the X$\alpha$ exchange energy and shows precisely how the LCAO KS potential and orbitals are obtained variationally from a single energy that depends on element-specific $\alpha$ values. The third section compares molecular dipole moments of a standard set of molecules computed with the flavors of ADFT considered so far. A problem is noted and a higher barrier to LCAO basis-set coalescence is needed to restrict the hydrogen orbital principally for the hydrogen atom if the spread of $\alpha$ values it too great. The fourth section considers the effect of different $\alpha$ values in the head and tail groups of $NH_2(CH)_{24}NO_2$ on its large dipole moment and compares with another calculation [19]. The final section contains concluding remarks.

## II. Variational fitting

One kind of electron-electron interaction that can be treated with ADFT is the self Coulomb repulsion of the electronic density,

$$E_C\left[\rho\right]=\tfrac{1}{2}\langle\rho|\rho\rangle\,, \tag{2}$$

where the Coulomb interaction, $1/r_{12}$, is indicated by the vertical line. As the density is the sum of the magnitude squared of occupied orbitals, this term can be treated directly in an $N^4$ LCAO method. Alternatively, given any approximation, $c(\mathbf{r})$, to $\rho(\mathbf{r})$ a unique approximation to $E_C$ [8],

$$E_C^{'}\left[\rho,c\right]=\langle\rho|c\rangle-\tfrac{1}{2}\langle c|c\rangle\,, \tag{3}$$

contains no first-order error, relative to Eq. (2), due to using the fit, $c$, to construct the approximate energy. An energy that is free of first-order errors is said to be robust [20]. Only robust energies can be made stationary, in this case with respect to variation, $\delta c(\mathbf{r})$, of the value of $c(\mathbf{r})$ at any position $\mathbf{r}$. This energy is stationary if

$$\langle\delta c|\rho-c\rangle=0\,, \tag{4}$$

for all variations. Thus $c = \rho$ everywhere, and in any fully variational context there is no difference between Eqs. (2) and (3). A robust fitted energy and the energy itself are identical if the fit is obtained variationally and the fitting basis set allows independent variations at every point in space.

It is also possible to treat Slater's Xα method analytically. A robust fitted approximation to the integral of the four-thirds power of the density is expressed [21],

$$E_x^{'}[\rho, x, y] = -\kappa \left\langle \tfrac{4}{3}\rho x - \tfrac{2}{3}xxy + \tfrac{1}{3}yy \right\rangle. \tag{5}$$

The calculus of variations with respect to all but the density yields,

$$\left\langle \tfrac{4}{3}(\rho - yx)\delta x - \tfrac{2}{3}(xx - y)\delta y \right\rangle = 0 \tag{6}$$

If all variations are allowed then $\rho = xy$ and $x^2 = y$. Thus $x = \sqrt[3]{\rho}$, and through the calculus of variations Slater's exchange energy is obtained precisely,

$$E_x^{'}[\rho, x, y] = -\kappa \left\langle \tfrac{4}{3}\rho x - \tfrac{2}{3}xxy + \tfrac{1}{3}yy \right\rangle = -\kappa \left\langle \rho^{4/3} \right\rangle = E_x[\rho]. \tag{7}$$

Note, the solution from the calculus of variations is unique, were we to solve $x^3(\mathbf{r}) = \rho(\mathbf{r})$ algebraically, then we would have obtained three solutions, $x(\mathbf{r}) = \exp(2n\pi i/3)\rho^{1/3}(\mathbf{r})$.

In the LCAO approach the orbitals, $c$, $x$, and $y$ are all LCAO expressions. If finite basis sets are used, then full variation freedom is not allowed and Eqs. (3) and (5) are, respectively, approximate expressions for the Coulomb and exchange energy. Eq. (5), multiplied by 2/3, has been used to obtain perfectly smooth ADFT potential energy surfaces using Gaussian basis sets [22].

Different uniform α values in the Xα method could be handled by multiplying Eq. (5), overall, by another α or equivalently $\rho$ in Eq. (5) by $\alpha^{3/4}$. In the following, the character α without superscript will be used to mean either or both of those possibilities and a function that can act on the LCAO density. An operator expression is needed when α is divided according to the centers of both of the orbital basis functions that appear in each term of the LCAO density. Even in that case, the calculus of variations ensures that $x$ is approximately the cube root of the appropriate quantity. As atoms separate the product of orbitals on different centers dies off as rapidly as the product of the two most diffuse Gaussians on the two centers. At infinite separation the molecular energies become the sum of the atomic energies given only by the α value that we choose for the subset of two-center densities that share a common atomic center. In a recent ADFT realization each orbital basis function in the density is multiplied by the 3/8 power of α chosen for the corresponding element [13]. Then each nonzero two-center density with different atomic centers has α value corresponding to the geometric average of the two atomic values for each center. At large separations the two, different-center LCAO density components are zero, but there can be still long-range forces due to charge transfer, rarely if ever in a spin polarized calculation, or static dipoles on each atom, which can happen always.

The ADFT total electronic energy is,

$$E[\rho, c, x, y] = \sum_i n_i \left\langle \phi_i^* \hat{h} \phi_i \right\rangle + E_C^{'}[\rho, c] + E_x^{'}[\alpha\rho, x, y] \quad (8)$$

where $n_i$ is the occupation number of orbital $\phi_i(\mathbf{r})$ and $\hat{h}$ is the one-electron part of the electronic Hamiltonian. In the following $c$, $x$, and $y$ are single-center LCAO expressions and $\rho$ is the two-center LCAO expression. Note that α multiplying Eq. (1) corresponds to $\alpha^{3/4}$ in this expression.

The derivative of the energy with respect to orbital coefficients, $C_\mu^i$, in the LCAO expression for each molecular orbital,

$$\phi_i(\mathbf{r}) = C_\mu^i u_\mu(\mathbf{r}) \tag{9}$$

where repeated indices are to be summed, yields the analytic Fock matrix,

$$(h_{\mu\nu} + \mathbf{c}\cdot\mathbf{C}_{\mu\nu} - \mathbf{x}\cdot\mathbf{X}_{\mu\nu})C_\mu^{i*}, \tag{10}$$

in the orbital basis with **c** and **x**, respectively, the LCAO fitting coefficients for the charge density and exchange-correlation part of the KS potential. **C** and **X** are the corresponding three-center Coulomb

$$C_{ij} = \left\langle c \middle| \phi_i^* \phi_j \right\rangle, \tag{11}$$

and overlap, exchange matrix elements,

$$X_{ij} = 4\kappa \left\langle \alpha x \phi_i^* \phi_j \right\rangle / 3, \tag{12}$$

which includes the factor $4\kappa\alpha/3$ to make Eq. (10) more compact. Note that at each step in the SCF process and at convergence the molecular orbital eigenvalues and eigenvectors are completely determined by **c** and **x** through the eigenvalue problem, which adds to Eq. (10) the Lagrange multipliers ensuring orbital orthonormality.

Variation of the energy with respect to charge-density fitting coefficients results in a set of equations,

$$\left\langle \rho \middle| c_i \right\rangle = \left\langle c_i \middle| c_j \right\rangle c_j, \tag{13}$$

where a subscripted quantity outside an integral indicates an LCAO fit coefficient, however inside an integral a subscripted quantity indicates the basis function, can be solved for the fit [8],

$$c_i = \left\langle \mathbf{c} \middle| \mathbf{c} \right\rangle_{ij}^{-1} \left\langle c_j \middle| \rho \right\rangle. \tag{14}$$

Optionally, a Lagrange multiplier can be used to constrain the fit to have the proper amount of charge, $\langle c \rangle = \langle \rho \rangle$. Taking the variation of the energy with respect to coefficients of *y* and solving gives that fit,

$$y_i = \langle \mathbf{yy} \rangle_{ij}^{-1} \cdot \langle y_j xx \rangle. \tag{15}$$

Variation of the energy with respect to coefficients of *x* and using the above equation results in a set of equations,

$$\langle \alpha\rho x_i \rangle = \langle x_i x y_j \rangle \langle \mathbf{yy} \rangle_{jk}^{-1} \langle y_k xx \rangle, \tag{16}$$

which can be solved by the time the SCF process is complete using only a single Newton-Raphson step each SCF cycle, with or without the constraint of another charge-conservation-like equality $\langle xy \rangle = \langle \alpha\rho \rangle$. All the calculations reported in this work use the two constraints. This equation also simplifies Eq. (5),

$$E_x^{''}[\alpha\rho, x] = -\kappa[\langle \tfrac{4}{3}\alpha\rho x \rangle - \tfrac{1}{3}\langle xx\mathbf{y} \rangle \cdot \langle \mathbf{yy} \rangle^{-1} \cdot \langle \mathbf{y}xx \rangle]. \tag{17}$$

Using Schwarz' historical HF α values [23] and standard orbital and fitting bases this gives a mean absolute error (MAE) for atomization energies of a standard set of 56 molecules [24] ranging from 15 to 20 Kcal/mole using various fitting an orbital basis sets [13]. Alternatively one can choose α to give the exact atomic (EA) energies for each element and basis-set combination. This gives a MAE on the *total* energy of these molecules that is a good as the best hybrid density-functional methods [15]. Because the α values are adjusted to get the exact atomic energies for each basis set the spread in MAE of *atomization* energies is much smaller16-17 Kcal/mol. If the α's do not vary too much from element to element as they do if, for example, they are chosen to given exact homonuclear diatomic molecule binding energies [25], then ADFT is computationally stable and reliable for energies. Thus, it is appropriate to look at another property, the dipole moment.

## III. Dipole Moments of the Extended G2 Set of Molecules

Symmetry is used to compute molecular dipole moments,

$$\mu_i = \left\langle \phi_j^* r_i \phi_j \right\rangle \tag{18}$$

as efficiently as the energy [26] and its gradient [27]. If $a_{i\lambda\mu}(\mathbf{r})$ is a component on atom $i$ of a symmetry-adapted function transforming as irreducible representation (irrep) $\lambda$ and row $\mu$ and $b_{j\lambda\mu}(\mathbf{r})$ is a component on atom $j$ of the same or another symmetry-adapted function transform as the same row of the same irrep, then an invariant is formed by summing over all rows [28],

$$\sum_\mu a_{i\lambda\mu}(\mathbf{r})\, b_{j\lambda\mu}(\mathbf{r}) \tag{19}$$

This theorem means that self-consistent-field (SCF) integrals, derivatives and dipole moments need only be evaluated for symmetry-distinct bra-ket atomic pairs of LCAO basis functions. For the small molecules considered in this work the dipole moment is computed extremely rapidly even without this efficiency. The dipole moments are computed in the Cartesian basis [29] and transformed by a matrix multiply [30] to the solid spherical harmonic basis of the ADFT calculation. There are only two orbital basis indices in Eq. (18), so the computer time need to compute the dipole moment in any manner is insignificant.

Fitting is well established in quantum chemistry and there are many choices for orbital bases and several choices for fitting basis sets. A package, A2, of basis functions for $c(\mathbf{r})$ and $x(\mathbf{r})$ (for fitting the VWN local-density functional [31]) was optimized [32] for use in DGauss [33] with a valence DZP orbital basis set DZVP2. We use this DZP orbital basis and 6-311G** [34,35], which is a TZP basis. Another valence TZP orbital basis set and a matched Resolution-of-the-Identity-J (RIJ) basis for $c(\mathbf{r})$ is optimized for use with the Turbomole program [36].

These basis sets, like all LCAO basis sets, are incomplete. Instead we rely on the variational principle and the stationary energy that it gives. Fitting basis functions of nonzero angular momentum are less important for fitting basis sets than they are for orbital basis sets. In the later basis they are essential to fit orbitals of nonzero orbital angular momentum. The density and thus the fits used to construct the KS potential are as a first approximation overlapping *s*-type, atomic functions [7,10,37]. It is important to get good *s*-type fitting basis sets. It is safest, and perhaps best, to scale the primative *s*-type orbital exponents [8] by 2, 2/3, and 4/3 to get the *s*-type basis for, respectively, the *c*, *x*, and *y* bases. This is correct in the single-zeta, separated, one-*s*-electron-atom limit. To supplement all scaled *s*-type basis we choose to use the RIJ non-*s* basis in all fitting bases used with the 6-311G** orbital basis because it has one unit of angular momentum more than the alternative A2 basis. This combination is called TZP/RIJ. We choose to use the A2 non-*s* *c* basis for all fitting bases used with the DZP orbital basis. This smaller combination is called DZP/A2. A larger set of basis set combinations has been tested [13]. Varying the orbital basis set has a much larger effect than any variation of fitting basis sets.

The dipole moment computed by Eq. (18) from orbitals and the KS potential that give a stationary energy is not necessarily stationary and therefore potentially less reliable than the total energy. It is important to examine the dependence of dipole moments on ADFT models. This work considers variation over the subset of molecules from the extended G2 set [38], whose symmetry allows a nonzero dipole moment. They are listed in Table I, along with their symmetries. Some of these symmetries, including those of CH, OH, ClO, HS, and $ClF_3$, which were mistakenly thought to have integral occupation numbers in higher symmetry, are different than as listed in our previous work [39]. Some molecules, including $N_2H_4$, $CH_3ONO$ and $(CH_4)_2SO$ had different shape than those found by Curtiss, et al. (http://chemistry.anl.gov/compmat/g2-97.htm), and have been corrected. These differences are Jahn-Teller-like energy differences and do not significantly lower the variational energies and thus do not significantly affect the conclusion of our previous work, but they do significantly affect the dipole moments. We say Jahn-Teller-like in this context because atoms and diatomic molecules cannot distort to remove degeneracy, but their ADFT density will distort to remove degeneracy unless constrained to be symmetric [26,40].

The third column in Table I gives the TZP (6-311G**) B3LYP [41,42] dipole moments computed using Gaussian03 [43]. All dipole moments are given at the optimized geometry for each basis set and method. All ADFT calculations start at one molecular geometry and superposed spherically symmetric atomic KS potentials from which the calculations converged to the desired structure. The listed molecular symmetry is used in all ADFT calculations. The ADFT results are given in the last three columns. HF $\alpha$ values are those that give the numerical HF energy for each element [23]. Exact atomic (EA) $\alpha$ values give experimental atomic energies when used in ADFT atomic calculations [15]. Optimized (Opt) $\alpha$ values minimize the MAE DZP/A2 error in atomization energies of the extended G2 set of molecules [39].

Table II lists the error in the magnitude of the dipole moments using different fits for exchange-only DFT. The dipole moment is a vector quantity. We compare magnitudes of the dipole moments. For small dipole moments we will underestimate the error of dipole moments that change sign or direction, but this is not a concern for the large differences that dominate our measures of correlation. The first column gives the difference between an $N^4$ calculation and a calculation using the DGA1 charge density fitting basis (with DensityFit keyword) and the TZP orbital basis of Gaussian03. These calculations used the default fine grid (pruned 75, 302 grid) with about 7000 points per atom. Using the larger ultrafine grid (99, 590 grid) the dipole moments change by at most 0.0002 Debye. Thus they are converged with respect to numerical grid size for purposes of the following comparisons. The second column gives the difference between the $N^4$ calculation and the ADFT calculation using the TZP/RIJ basis combination. The mean error and mean absolute error (MAE) are quite small for both $N^3$ calculations. The MAE's are less than 0.1 Debye. On average the dipole moments obtained using the density fitting option are slightly smaller those obtained with direct $N^4$ calculation. The ADFT errors are twice as big as the DensityFit errors. The third column uses the smaller DZP/A2 basis combination. The size of the dipole moment error and the name of molecules of the extreme outliers are listed in the final four rows of the table. The molecules that most over-estimate dipole moments have only first-row atoms. Those that most underestimate dipole moments have second-row atoms. Both ADFT calculations have the same molecule ClNO with the largest too-small fitted dipole moment. Most likely these errors are systematic and can be improved, but even at this stage of development ADFT dipole moments are acceptable. The DZP/A2 basis combination considerably under-predicts the dipole of LiF.  LiF is the maximal outlier for many other ADFT models also when compared with experiment.  (See Table IV).

Table III assumes B3LYP/TZP to be a possible target for ADFT.  In the second column ADFT with exchange only, $\alpha$=2/3, and the TZP/RIJ basis combination is compared to this target.  The third column uses the same methodology, but with $\alpha = 0.7$ [4].  This is slightly different from the best uniform $\alpha$, 0.7065 for basis TZP/RIJ and 0.698 for basis DZP/A2 obtained by minimizing the mean absolute error (MAE) of atomization energies with respect to this parameter over the G2 set of molecules [14]. Judging from mean error and MAE 0.7 is a better choice of $\alpha$ value than 2/3.  On the other hand, judging from the extreme outliers 2/3 is the better choice.  The third column uses the historical $\alpha$ values, which give the HF energies in numerical calculations [23]. The fourth column uses the $\alpha$ values that give exact atomic energies for the TZP/RIJ basis combination [15].  The fifth column uses the $\alpha$ values that with the DZP/A2 basis combination minimizes the MAE of atomization energies over the G2 set of molecules [39].  All measures: mean error, mean absolute error, maximum error and minimum error indicate that B3LYP is closest to the uniform $\alpha$ ADFT models.  Again all molecules that have the largest errors include second-row atoms.  In this case they all include the highest atomic-weight atom chlorine for which gradients of the density, and perhaps gradient corrections in B3LYP, are the largest.

The optimum $\alpha$ values of the last column of Table III are absolutely well defined but are also a bit problematical.  The most over-estimated dipole moment is for BeH. There is only one molecule in the extended G2 set that has Be.  So given an optimal $\alpha$ value for the set of hydrogen-containing atoms, the beryllium $\alpha$ value will be the value that exactly gives the BeH bond energy. A better choice would be to split the difference and miss the dipole moment less and the atomization energy more. Expanding our database as we do with this work and reoptimizing the elemental $\alpha$ values would make them more uniform. That is essential in our model that uses the geometrical mean of alpha values. These values of $\alpha$ cannot be used with the TZP/RIJ basis combination. For calculations on $NH_3$, $CH_3NH_2$, ethylamine, and $(CH_3)_2CH$ it is straight down hill for an H atom to fall into a C or N atom to a distance of 0.6 Angstroms at which point the SCF process dies. In this flavor ADFT wants to use the H orbital as a heavy-atom $sp^3$ orbital thereby lowering energy due to a higher $\alpha$ value. This same coalescence was seen in an Opt $\alpha$ DZP/A2 calculation in molecular-dynamics simulations of vibrationally excited PETN collisions [44]. Superposed basis sets are always a problem in the LCAO method and atoms must stay separated, *e.g.,* because uncontracted valence orbitals of each atom typically span the same space if the basis sets are multiple-zeta and balanced. Most of the time LCAO basis-set-superposition errors are small and decrease with the size of the basis set. We expect, but have not exhaustively verified, that that is the case in ADFT when atoms are separated by chemical and further distances. In any event, in ADFT this problem is potentially most significant only for H, which has the largest $\alpha$ value and no core electrons to independently provide separation.

This problem can be lessened without further limiting the differences in atomic $\alpha$ values that are allowed. A barrier that prevents coalescence in ADFT can be achieved simply by choosing the two-center $\alpha$ value to be the minimum of both atomic values. Such a change will benefit EA calculations. They overestimate the atomization energy of the extended G2 set with a mean error of 25 Kcal/mole [39]. Lowering the two-center $\alpha$ values will reduce the molecular energy without affecting at all the one-center terms in the LCAO density that exclusively determine the atomic energies. Thus increasing the barrier to basis-set coalescence will also lower the EA mean error. The code has been changed, but this work reports on ADFT calculations as previously defined.

Table IV is like Table III but takes experiment as the target for ADFT. The experimental data is from Ref. [45]. Only 83 of the 107 molecules of Table I have experimental dipole moments listed in that book, so our averages are over that subset. In this case the ADFT mean and mean absolute errors are quite good. For the first time an ADFT model, $\alpha$ = 0.7, on average overestimates the target dipole moments. The worst comparison with experiment is for the Opt $\alpha$ model, this data favors none of the other four models over another. This time, in contrast to the other comparison tables, the dipole moments of a chlorine-containing molecule have their dipole moments maximally overestimated and first-row-element molecule LiF has its dipole moment maximally underestimated by the four ADFT models that give better dipole moments in comparison with experiment. Compared to experiment the first four models are comparable. In contrast to Table III, which compares to B3LYP, all four MAE errors round to 0.2 D. Element-dependent exchange is a new, and perhaps useful, dimension in electronic structure calculations.

## IV. Dipole moment of $NH_2(CH)_{24}NO_2$

ADFT works well for molecular standards, but would appear to have a problem, however, with long “conducting” molecules. Changing the value of $\alpha$ of an atom in a molecule makes it give or take electrons to or from the rest of the molecule. There might be huge effects on the dipole moment of two atoms connected by a long conducting wire if there is a relative change in the two values of $\alpha$ at the ends of the molecule. Undoped trans-polyacetylene is not a metal because of Peierls dimerization, the magnitude of which is underestimated in DFT, likely because of the underestimation of the HOMO-LUMO gap in DFT [46]. Similarly one would expect polarization to be overestimated in DFT [29]. Such an analysis is consistent with calculations [19] on the push-pull $\pi$-conjugated molecule $NH_2(CH)_{24}NO_2$ which show that DFT calculations greatly overestimate the dipole moment, polarizability, and hyperpolarizabilities of this molecule relative to HF-based MP2. The dipole moments from these calculations are given in the first two rows of Table V. The basis sets are not polarized and the $\alpha = 2/3$ calculation is performed at the MP2 geometry. The DFT dipole moment is two and a half times as large as the MP2 dipole moment. This spectacular difference is perhaps even underestimated because the dimerization that one would find in a DFT optimization is expected to be less than the MP2 dimerization. The use of exact exchange via an optimized effective potential in a DFT method does correct this problem [47], but this work is an assessment of ADFT, not DFT.

The ADFT dipole moments for $NH_2(CH)_{24}NO_2$ are given in the last six rows of the Table. The moments are calculated at the optimized geometry of each model. The relevant parameter for testing whether or not variable $\alpha$ values in ADFT forces large transfers of electrons is the difference between the values on the two hydrogen atoms on one end of the molecule and values on the two oxygen atoms at the other end of the molecule. That different is listed in the third row of the table. Remarkably there is essentially no effect. The TZP moments are about ten percent smaller than the DZP moments. Within each basis-set combination the middle difference in $\alpha$ values (Exact Atomic) yields the largest dipole moment.

## V. Conclusions

We have examined the performance of ADFT for predicting molecular dipole moments. The MAE in the dipole moment of the extended G2 set of molecules simply due to fitting the KS potential is less than 0.1 Debye. B3LYP dipole moments most closely match the dipole moments from the uniform-exchange ADFT models. That is not the case for our comparison with experimental dipole moments. Thus the extra dimension of element-specific exchange in ADFT may be generally useful in quantum chemistry.

The dipole moment of $NH_2(CH)_{24}NO_2$ is largely unaffected by different non-uniform $\alpha$ ADFT models. Different values of $\alpha$ on the ends of the molecule do not affect the total dipole moment much, if at all. Changing $\alpha$ values in ADFT apparently only affects the charge distribution locally, but, of course, all local differences sum up to the global differences between the ADFT models corresponding to the differences between the columns of Table I.

ADFT combines some advantages of both conventional *ab initio* and conventional DFT calculations, but has limitations including, so far, the inability to compute gradient corrections. For energetics this limitation can be overcome through element-dependent $\alpha$ values. Variable $\alpha$ values do not significantly affect dipole moments. Similarly, DZP dipole moments are weakly dependent choice of density functional or correlation method for standard molecules [48]. ADFT seems a viable method for obtaining density-functional dipole moments of molecules with reasonable accuracy in a reasonable amount of computer time, which is important in developing an atomistic model of electrochemistry and can be extended to compute dipole polarizabilities for developing advanced nonlinear optical materials.

## Acknowledgment

The Office of Naval Research, directly and through the Naval Research Laboratory, and the DoD's High Performance Computing Modernization Program, through the Common High Performance Computing Software Support Initiative, Project MBD-5, supported this research. The Gaussian03 calculations were performed at the ARL MSRC.

Table I. Nonzero dipole moments (in Debye) computed for the extended G2 set of molecules [43] using various methods. The molecular symmetry is that used in the ADFT calculations the results of which are given in the last three columns. The B3LYP dipole moments are calculated using Gaussian03 [34]. HF $\alpha$ values are those that give the numerical HF energy for each element [23]. Exact atomic (EA) $\alpha$ values give experimental atomic energies when used in ADFT atomic calculations [15]. Optimized (Opt) $\alpha$ values minimize the MAE error in atomization energies of the extended G2 set of molecules [39]. The last column gives experimental values [45]. TZP and DZP indicate, respectively, calculations using the 6-311G** [34,35] and DZVP2 [32] orbital bases. RIJ is the Turbomole non-*s* fitting basis for *c* [36] and A2 is the DGauss non-*s* *c* fitting basis [32].

| Molecule | Symmetry | B3LYP | $\alpha$=2/3 | HF $\alpha$ | EA $\alpha$ | Opt $\alpha$ | Exp. |
|---|---|---|---|---|---|---|---|
| Bases | | TZP | TZP/RIJ | TZP/RIJ | TZP/RIJ | DZP/A2 | |
| LiH | $C_{6v}$ | 5.703 | 5.515 | 5.614 | 5.476 | 5.352 | 5.884 |
| BeH | $C_{6v}$ | 0.291 | 0.203 | 0.299 | 0.166 | 1.799 | |
| CH | $C_{2v}$ | 1.504 | 1.432 | 1.455 | 1.562 | 1.255 | 1.46 |
| $CH_2$(T) | $C_{2v}$ | 0.648 | 0.622 | 0.503 | 0.588 | 0.476 | |
| $CH_2$(S) | $C_{2v}$ | 1.924 | 1.887 | 1.899 | 2.019 | 1.711 | |
| NH | $C_{6v}$ | 1.599 | 1.548 | 1.517 | 1.604 | 1.245 | 1.39 |
| $NH_2$ | $C_{2v}$ | 1.962 | 1.959 | 1.908 | 2.000 | 1.543 | |
| $NH_3$ | $C_{3v}$ | 1.714 | 1.752 | 1.691 | 1.665 | 1.183 | 1.471 |
| OH | $C_{2v}$ | 1.760 | 1.761 | 1.689 | 1.771 | 1.591 | 1.668 |
| $H_2O$ | $C_{2v}$ | 2.070 | 2.099 | 2.005 | 2.088 | 1.716 | 1.854 |
| HF | $C_{6v}$ | 1.902 | 1.916 | 1.805 | 1.893 | 1.548 | 1.826 |
| LiF | $C_{6v}$ | 5.750 | 5.419 | 5.415 | 5.476 | 5.632 | 6.326 |
| CN | $C_{6v}$ | 1.303 | 1.185 | 1.433 | 1.516 | 0.999 | |
| HCN | $C_{6v}$ | 2.941 | 2.879 | 2.792 | 2.895 | 2.419 | 2.984 |
| CO | $C_{6v}$ | 0.121 | 0.261 | 0.285 | 0.275 | 0.323 | 0.11 |
| HCO | $C_{1h}$ | 1.604 | 1.414 | 1.420 | 1.531 | 1.301 | |
| $H_2CO$ | $C_{2v}$ | 2.241 | 1.995 | 1.964 | 2.104 | 2.030 | 2.332 |
| $CH_3OH$ | $C_{1h}$ | 1.715 | 1.621 | 1.544 | 1.631 | 1.449 | 1.7 |
| $N_2H_4$ | $C_2$ | 2.060 | 1.935 | 1.708 | 1.693 | 0.890 | 1.75 |
| NO | $C_{2v}$ | 0.076 | 0.173 | 0.176 | 0.166 | 0.070 | 0.159 |
| $SiH_2$(S) | $C_{2v}$ | 0.150 | 0.089 | 0.331 | 0.149 | 0.706 | |
| $SiH_2$(T) | $C_{2v}$ | 0.164 | 0.096 | 0.484 | 0.379 | 0.146 | |
| $PH_2$ | $C_{2v}$ | 0.714 | 0.729 | 0.332 | 0.477 | 1.883 | |
| $PH_3$ | $C_{3v}$ | 0.771 | 0.818 | 0.335 | 0.512 | 2.243 | 0.574 |
| $H_2S$ | $C_{2v}$ | 1.332 | 1.420 | 1.034 | 1.164 | 1.389 | 0.97 |
| HCl | $C_{6v}$ | 1.422 | 1.453 | 1.178 | 1.260 | 0.990 | 1.109 |
| NaCl | $C_{6v}$ | 9.101 | 8.521 | 8.531 | 8.510 | 6.671 | 9.001 |
| SiO | $C_{6v}$ | 2.891 | 2.542 | 2.833 | 2.922 | 1.855 | 3.098 |
| CS | $C_{6v}$ | 1.840 | 1.917 | 2.304 | 2.386 | 1.349 | 1.958 |
| SO | $C_{6v}$ | 1.711 | 1.606 | 1.857 | 1.929 | 1.108 | 1.55 |
| ClO | $C_{2v}$ | 1.388 | 1.616 | 1.856 | 1.915 | 1.682 | 1.239 |
| ClF | $C_{6v}$ | 1.229 | 1.152 | 1.256 | 1.294 | 0.678 | 0.888 |
| $CH_3Cl$ | $C_{3v}$ | 2.129 | 1.891 | 1.490 | 1.572 | 1.617 | 1.892 |
| $CH_3SH$ | $C_{1h}$ | 1.713 | 1.663 | 1.258 | 1.374 | 1.815 | 1.52 |
| HOCl | $C_{1h}$ | 1.785 | 1.755 | 1.637 | 1.714 | 1.479 | 1.3 |
| $SO_2$ | $C_{2v}$ | 1.956 | 1.820 | 1.961 | 2.001 | 1.363 | 1.633 |
| COS | $C_{6v}$ | 0.531 | 0.582 | 0.996 | 1.111 | 0.024 | 0.715 |
| $COF_2$ | $C_{2v}$ | 0.995 | 1.052 | 1.161 | 1.161 | 1.656 | 0.95 |
| $N_2O$ | $C_{6v}$ | 0.058 | 0.228 | 0.229 | 0.217 | 0.005 | 0.161 |
| $CF_3CN$ | $C_{3v}$ | 1.248 | 1.408 | 1.456 | 1.431 | 1.377 | 1.262 |
| $CH_3CCH$ | $C_{3v}$ | 0.784 | 0.855 | 0.877 | 0.882 | 1.075 | 0.784 |
| $C_3H_4$ | $C_{2v}$ | 0.499 | 0.525 | 0.511 | 0.495 | 0.420 | 0.45 |
| $CH_3CHCH_2$ | $C_{1h}$ | 0.381 | 0.416 | 0.445 | 0.446 | 0.443 | 0.366 |
| $C_3H_8$ | $C_{2v}$ | 0.077 | 0.091 | 0.063 | 0.077 | 0.060 | 0.084 |
| $C_4H_6$ | $C_{2v}$ | 0.419 | 0.473 | 0.431 | 0.472 | 0.407 | |
| bicyclobutane | $C_{2v}$ | 0.752 | 0.795 | 0.705 | 0.752 | 0.599 | |
| cyclobutene | $C_{2v}$ | 0.137 | 0.193 | 0.191 | 0.228 | 0.272 | 0.132 |
| isobutene | $C_{2v}$ | 0.531 | 0.556 | 0.620 | 0.609 | 0.675 | 0.503 |
| isobutane | $C_{3v}$ | 0.119 | 0.152 | 0.115 | 0.130 | 0.036 | 0.132 |
| $CH_2F_2$ | $C_{2v}$ | 1.934 | 1.697 | 1.549 | 1.684 | 1.243 | 1.978 |
| $CHF_3$ | $C_{3v}$ | 1.605 | 1.431 | 1.289 | 1.403 | 1.005 | 1.651 |
| $CH_2Cl_2$ | $C_{2v}$ | 1.859 | 1.669 | 1.256 | 1.341 | 1.364 | 1.6 |
| $CHCl_3$ | $C_{3v}$ | 1.243 | 1.097 | 0.761 | 0.830 | 0.835 | 1.04 |
| $CH_3NH_2$ | $C_{1h}$ | 1.379 | 1.335 | 1.236 | 1.252 | 0.903 | 1.31 |
| $CH_3CN$ | $C_{3v}$ | 3.880 | 3.870 | 3.777 | 3.906 | 3.530 | 3.924 |
| $CH_3NO_2$ | $C_{1h}$ | 3.517 | 3.214 | 3.156 | 3.299 | 3.227 | 3.46 |
| $CH_3ONO$ | $C_{1h}$ | 2.202 | 2.024 | 2.106 | 2.285 | 1.721 | |
| $CH_3SiH_3$ | $C_{3v}$ | 0.743 | 0.852 | 0.806 | 0.790 | 1.110 | 0.735 |
| HCOOH | $C_{1h}$ | 1.429 | 1.386 | 1.469 | 1.505 | 1.571 | 1.41 |
| $HCOOCH_3$ | $C_{1h}$ | 1.833 | 1.763 | 1.791 | 1.869 | 1.677 | 1.77 |
| $CH_3CONH_2$ | $C_{1h}$ | 3.753 | 3.655 | 3.636 | 3.756 | 3.288 | 3.76 |
| $C_2H_4NH$ | $C_{2v}$ | 1.727 | 1.664 | 1.559 | 1.653 | 1.334 | 1.9 |
| $(CH_3)_2NH$ | $C_{1h}$ | 0.978 | 0.866 | 0.810 | 0.845 | 0.600 | 1.01 |
| $CH_3CH_2NH_2$ | $C_{1h}$ | 1.273 | 1.507 | 1.380 | 1.390 | 0.718 | 1.22 |
| $CH_2CO$ | $C_{2v}$ | 1.423 | 1.364 | 1.252 | 1.421 | 1.038 | 1.442 |
| $C_2H_4O$ | $C_{2v}$ | 1.968 | 1.777 | 1.656 | 1.779 | 1.648 | 1.89 |
| $CH_3CHO$ | $C_{1h}$ | 2.662 | 2.362 | 2.364 | 2.508 | 2.520 | 2.75 |
| $CH_3CH_2OH$ | $C_{1h}$ | 1.605 | 1.531 | 1.447 | 1.533 | 1.307 | 1.44 |
| $(CH_3)_2O$ | $C_{2v}$ | 1.297 | 1.308 | 1.222 | 1.312 | 1.294 | 1.3 |
| $C_2H_4S$ | $C_{2v}$ | 2.140 | 1.988 | 1.428 | 1.536 | 2.180 | |
| $(CH_3)_2SO$ | $C_{1h}$ | 4.148 | 3.781 | 3.762 | 3.910 | 4.274 | 3.96 |
| $CH_3CH_2SH$ | $C_{1h}$ | 1.793 | 1.730 | 1.300 | 1.413 | 1.857 | 1.58 |
| $(CH_3)_2S$ | $C_{2v}$ | 1.712 | 1.749 | 1.380 | 1.492 | 2.016 | 1.554 |
| $CH_2CHF$ | $C_{1h}$ | 1.400 | 1.171 | 1.024 | 1.148 | 0.662 | 1.468 |
| $CH_3CH_2Cl$ | $C_{1h}$ | 2.345 | 2.085 | 1.632 | 1.717 | 1.746 | 2.05 |
| $CH_2CHCl$ | $C_{1h}$ | 1.662 | 1.313 | 0.808 | 0.886 | 0.919 | 1.45 |
| $CH_2CHCN$ | $C_{1h}$ | 3.904 | 3.800 | 3.686 | 3.835 | 3.347 | 3.87 |
| $(CH_3)_2CO$ | $C_{2v}$ | 2.845 | 2.476 | 2.475 | 2.614 | 2.699 | 2.88 |
| $CH_3COOH$ | $C_{1h}$ | 1.600 | 1.534 | 1.609 | 1.655 | 1.769 | 1.7 |
| $CH_3COF$ | $C_{1h}$ | 2.869 | 2.754 | 2.681 | 2.816 | 2.752 | 2.96 |
| $CH_3COCl$ | $C_{1h}$ | 3.037 | 2.802 | 2.504 | 2.626 | 2.517 | 2.72 |
| $CH_3CH_2CH_2Cl$ | $C_{1h}$ | 2.444 | 2.176 | 1.708 | 1.785 | 1.797 | 2.05 |
| $(CH_3)_2CHOH$ | $C_{1h}$ | 1.578 | 1.602 | 1.502 | 1.586 | 1.386 | 1.58 |
| $CH_3CH_2OCH_3$ | $C_{1h}$ | 1.194 | 1.145 | 1.065 | 1.162 | 1.005 | 1.17 |
| $(CH_3)_3N$ | $C_{3v}$ | 0.495 | 0.246 | 0.272 | 0.302 | 0.767 | 0.612 |
| $C_4H_4O$ | $C_{2v}$ | 0.621 | 0.488 | 0.384 | 0.469 | 0.370 | 0.66 |
| $C_4H_4S$ | $C_{2v}$ | 0.559 | 0.363 | 0.097 | 0.010 | 0.593 | 0.55 |
| $C_4H_4NH$ | $C_{2v}$ | 1.942 | 2.042 | 2.021 | 2.000 | 1.696 | 1.74 |
| $C_5H_5N$ | $C_{2v}$ | 2.190 | 2.143 | 2.074 | 2.171 | 2.047 | 2.215 |
| HS | $C_{2v}$ | 1.026 | 1.062 | 0.792 | 0.883 | 1.047 | 0.758 |
| CCH | $C_{6v}$ | 0.767 | 0.748 | 0.557 | 0.646 | 0.462 | |
| $CH_2CH$ | $C_{1h}$ | 0.678 | 0.698 | 0.595 | 0.676 | 0.599 | |
| $CH_3CO$ | $C_{1h}$ | 2.420 | 2.286 | 2.280 | 2.434 | 2.312 | |
| $CH_2OH$ | $C_1$ | 1.547 | 1.571 | 1.504 | 1.558 | 1.400 | |
| ClNO | $C_{1h}$ | 2.583 | 2.129 | 1.352 | 1.128 | 1.814 | |
| $NF_3$ | $C_{3v}$ | 0.189 | 0.080 | 0.026 | 0.027 | 0.085 | 0.235 |
| $PF_3$ | $C_{3v}$ | 1.552 | 1.364 | 1.442 | 1.497 | 1.513 | 1.03 |
| $O_3$ | $C_{2v}$ | 0.618 | 0.571 | 0.588 | 0.593 | 0.625 | 0.534 |
| $F_2O$ | $C_{2v}$ | 0.301 | 0.214 | 0.155 | 0.155 | 0.102 | 0.297 |
| $ClF_3$ | $C_{2v}$ | 1.015 | 0.980 | 1.069 | 1.098 | 0.585 | 0.6 |
| $CH_3O$ | $C_{1h}$ | 2.048 | 2.113 | 2.035 | 2.182 | 2.020 | |
| $CH_3CH_2O$ | $C_1$ | 2.023 | 2.174 | 2.086 | 2.217 | 2.105 | |
| $CH_3S$ | $C_{1h}$ | 1.798 | 1.815 | 1.425 | 1.533 | 2.038 | |
| $CH_3CH_2$ | $C_{1h}$ | 0.277 | 0.338 | 0.350 | 0.348 | 0.451 | |
| $(CH_3)_2CH$ | $C_{1h}$ | 0.193 | 0.218 | 0.241 | 0.250 | 0.380 | |
| $(CH_3)_3C$ | $C_{3v}$ | 0.192 | 0.164 | 0.202 | 0.207 | 0.380 | |
| $NO_2$ | $C_{2v}$ | 0.313 | 0.225 | 0.264 | 0.283 | 0.335 | 0.316 |

Table II. Error in dipole moments in Debye due to fitting the KS potential compared to an $N^4$ calculation. DF indicates the DensityFit option of Gaussian03. MAE is mean absolute error. For the two basis-set combinations see Table I. Under the maximum and minimum errors are listed the molecules for which those errors occurred.

| Basis | TZP/DF | TZP/RIJ | DZP/A2 |
|---|---|---|---|
| Mean Error | 0.011 | 0.027 | -0.116 |
| MAE | 0.04 | 0.070 | 0.175 |
| Minimum Error | -0.337 | -0.296 | -1.318 |
| Molecule | $CH_3CH_2$ | $CH_3CH_2NH_2$ | LiF |
| Maximum Error | 0.3 | 0.362 | 0.829 |
| Molecule | $SiH_2S$ | ClNO | ClNO |

Table III. The mean and mean absolute errors in the dipole moments (in Debye) calculated in various $N^3$ ADFT models relative to $N^4$ TZP B3LYP dipole moments. See Table II for definition of basis. The first two ADFT model have uniform $\alpha$ values and the last three have element-dependent $\alpha$ values.

| ADFT Model | $\alpha$=2/3 | $\alpha$=0.7 | HF $\alpha$ | EA $\alpha$ | Opt $\alpha$ |
|---|---|---|---|---|---|
| Basis | TZP/RIJ | TZP/RIJ | TZP/RIJ | TZP/RIJ | DZP/A2 |
| Mean Error | 0.070 | 0.055 | 0.145 | 0.085 | 0.177 |
| MAE | 0.117 | 0.109 | 0.218 | 0.177 | 0.349 |
| Minimum Error | -0.235 | -0.287 | -0.468 | -0.580 | -1.508 |
| Molecule | $CH_3CH_2NH_2$ | ClO | ClO | COS | BeH |
| Maximum Error | 0.579 | 0.687 | 1.231 | 1.454 | 2.430 |
| Molecule | NaCl | ClNO | ClNO | ClNO | NaCl |

Table IV. Difference relative to experimental values of dipole moments (in Debye) calculated in various $N^3$ ADFT models. Same as Table III, but with a different target and set of molecules.

| ADFT Model | $\alpha$=2/3 | $\alpha$=0.7 | HF $\alpha$ | EA $\alpha$ | Opt $\alpha$ |
|---|---|---|---|---|---|
| Basis | TZP/RIJ | TZP/RIJ | TZP/RIJ | TZP/RIJ | DZP/A2 |
| Mean Error | 0.006 | -0.010 | 0.081 | 0.018 | 0.164 |
| MAE | 0.175 | 0.162 | 0.223 | 0.194 | 0.344 |
| Minimum Error | -0.455 | -0.459 | -0.617 | -0.676 | -1.669 |
| Molecule | HOCL | HOCL | ClO | ClO | $PH_3$ |
| Maximum Error | 0.907 | 0.832 | 0.911 | 0.85 | 2.33 |
| Molecule | LiF | LiF | LiF | LiF | NaCl |

Table V. Dipole moment of $NH_2(CH)_{24}NO_2$ with various models and basis sets described in the text.

| Model | Basis | $\alpha_H - \alpha_O$ | $\mu$ (Debye) |
|---|---|---|---|
| $\alpha$=2/3 | 6-31G (DZ) | | 25.0 |
| MP2 | 6-31G (DZ) | | 10.6 |
| $\alpha$=2/3 | DZP | 0.000 | 30.1 |
| Hartree-Fock $\alpha$ | DZP | 0.034 | 29.5 |
| Exact Atomic $\alpha$ | DZP | 0.012 | 30.2 |
| $\alpha$=2/3 | TZP | 0.000 | 27.9 |
| Hartree-Fock $\alpha$ | TZP | 0.034 | 27.5 |
| Exact Atomic $\alpha$ | TZP | 0.012 | 28.3 |